 \documentclass[final,5p,times]{elsarticle}

 \usepackage{graphics}

\usepackage{amssymb}

\usepackage{subfig}
\usepackage{color}

\journal{NIM A  RICAP-2013}

\begin{document}

\begin{frontmatter}



\title{ Observation of TeV-PeV cosmic ray anisotropy with IceCube, IceTop and AMANDA }


\author{Paolo Desiati, for the IceCube Collaboration$^{\dagger}$}

\address{Wisconsin IceCube Particle Astrophysics Center (WIPAC) and Department of Astronomy University of Wisconsin, Madison, WI 53706, U.S.A.\\$\dagger$~http://icecube.wisc.edu/collaboration/authors/current}

\begin{abstract}
The study of cosmic ray anisotropy in the TeV-PeV energy range could provide clues about the origin and propagation of cosmic rays in our galactic neighborhood. Because the observed anisotropy is very small, below the per-mille level, large event volumes are needed in order to characterize it in sufficient detail.
Over the last six years, the IceCube Observatory has collected 150 billion cosmic ray induced muon events. This large data sample made it possible to detect anisotropies in the southern hemisphere, down to the 10$^{-5}$ level, at primary energies in excess of 10 TeV. The observed anisotropy is not a simple dipole, but it can be described as composed of multipole components of the spherical harmonic expansion, to about 10$^{\circ}$. A change in topological structure of the cosmic ray arrival distribution is observed above 100 TeV. Data collected with the air shower array IceTop above 300 TeV confirm the observations up to the PeV energy scale. Moreover, the addition of data collected with the AMANDA neutrino telescope, which operated between 2000 and 2007, has enabled us to search for time variability in the observed TeV anisotropy.
\end{abstract}


\end{frontmatter}



\section{Introduction}
\label{sec:intro}

\begin{figure*}[t]
\begin{tabular}{cc}
\subfloat[IceCube {\bf 20 TeV}]{\includegraphics[width=0.45\textwidth]{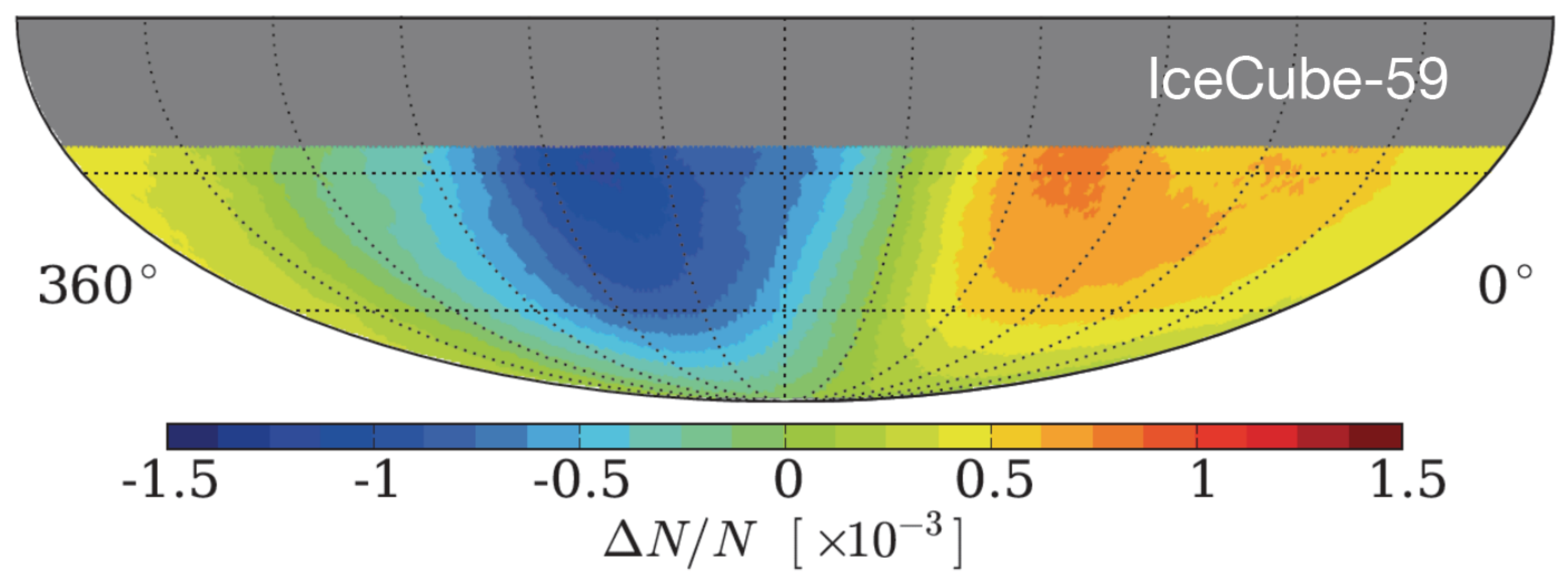} \label{fig:1mapa}} & 
\subfloat[IceCube {\bf 400 TeV}]{\includegraphics[width=0.45\textwidth]{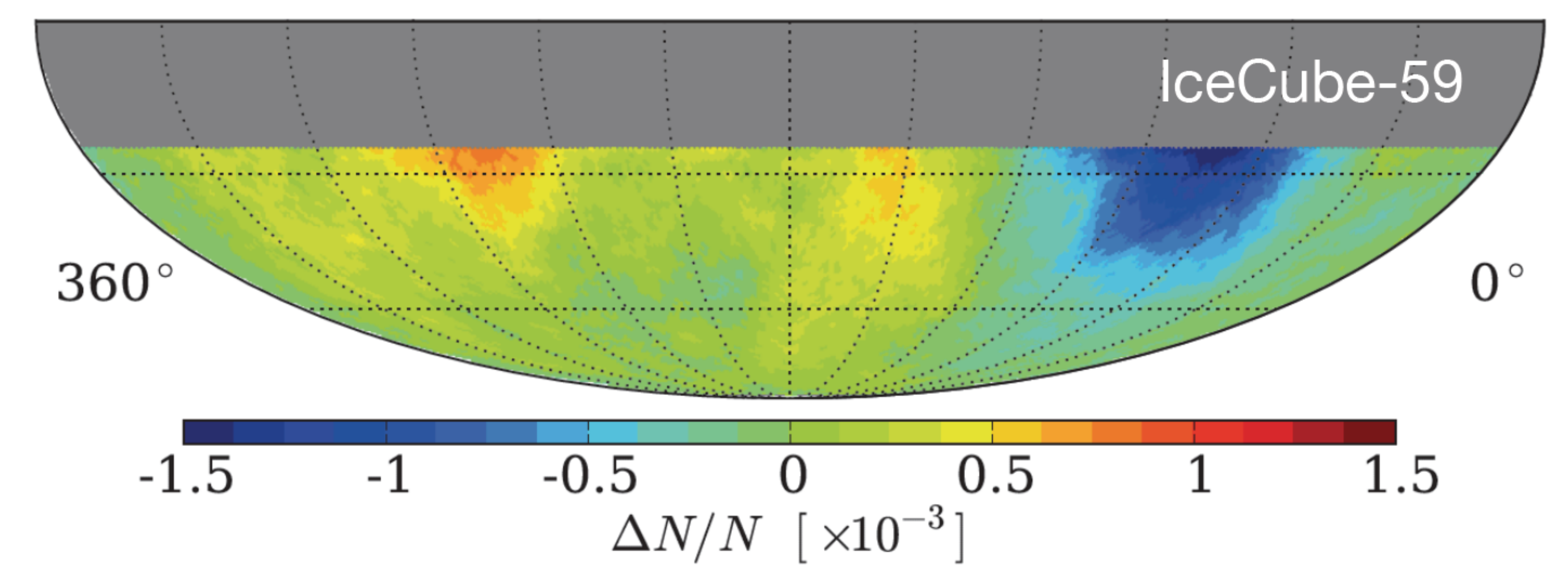} \label{fig:1mapb}} \\
\multicolumn{2}{r}{ } \\
\subfloat[IceTop {\bf 400 TeV}]{\includegraphics[width=0.45\textwidth]{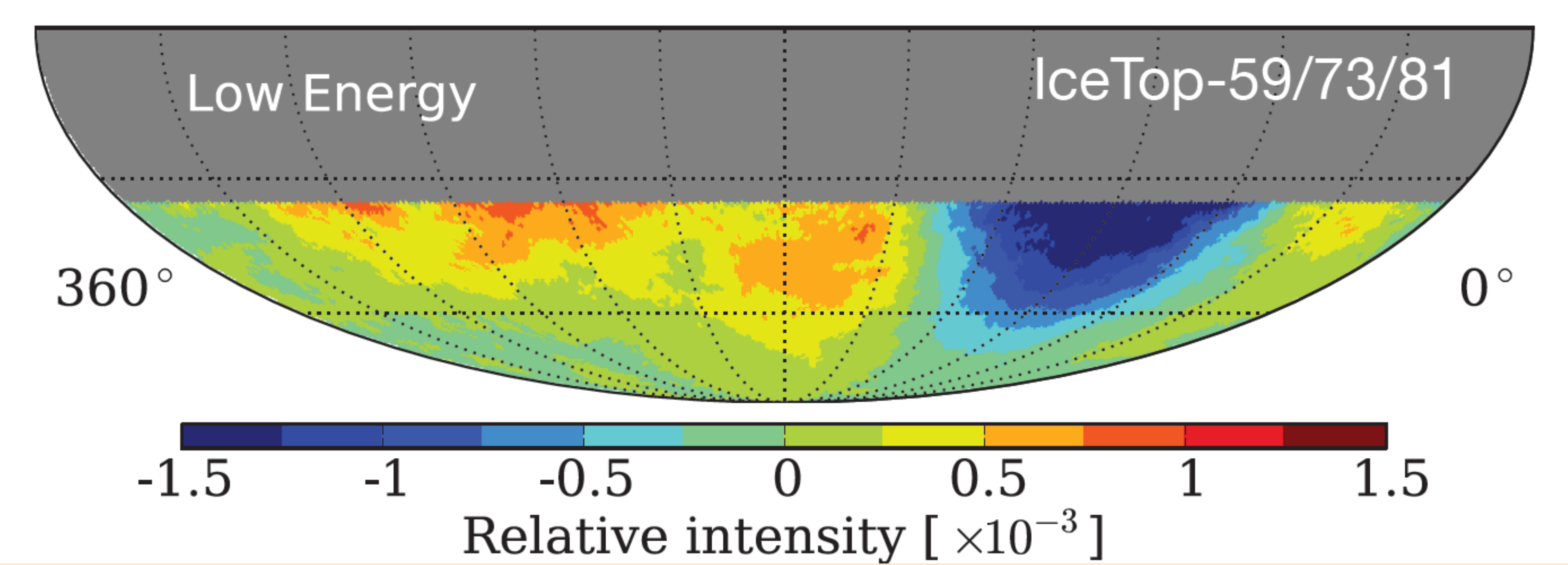} \label{fig:1mapc}} &
\subfloat[IceTop {\bf 2 PeV}]{\includegraphics[width=0.45\textwidth]{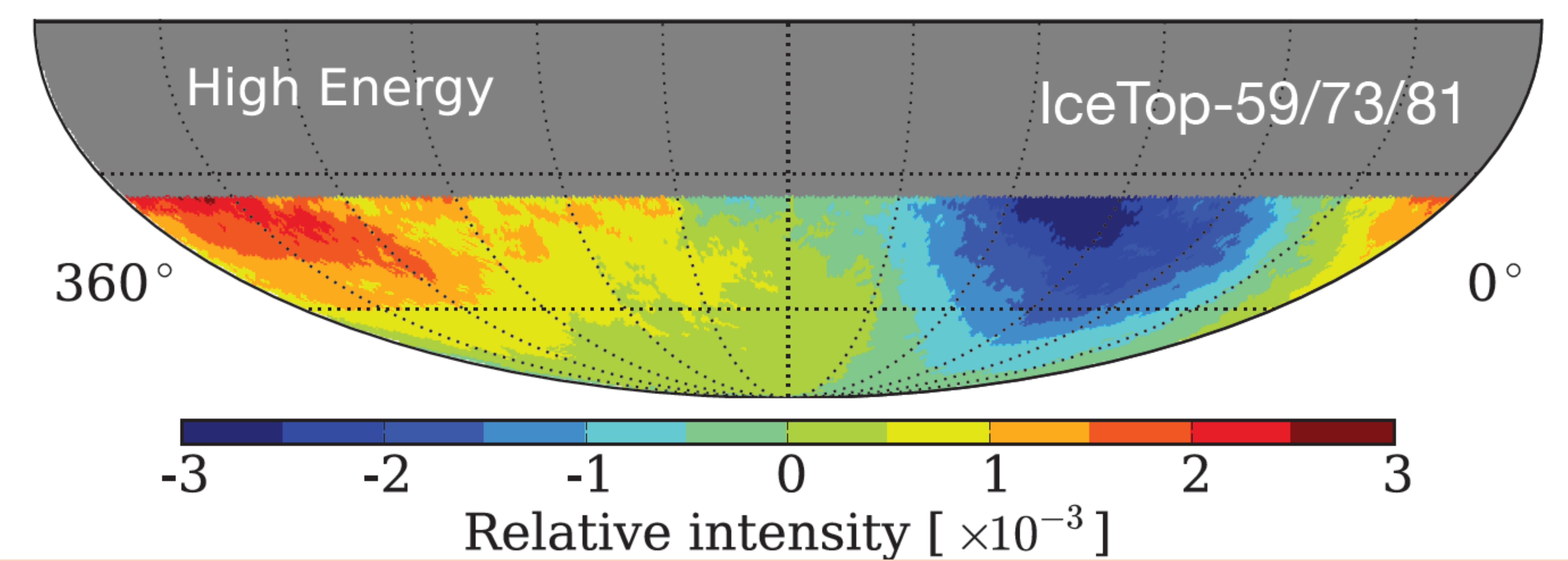} \label{fig:1mapd}} \\ 
\end{tabular}
\caption{2D maps of relative intensity in equatorial coordinates of the cosmic ray arrival distribution for IceCube (at 20 TeV and 400 TeV median energies)~\cite{iclarge12} and IceTop (at 400 TeV and 2 PeV median energies)~\cite{itlarge13}.}
\label{fig:1maps}
\end{figure*}

Over a century after their discovery, the origin of cosmic rays is still unknown. The current leading model is that cosmic rays are accelerated in diffusive shocks. In this case, Supernova Remnants (SNRs) in our Galaxy could be the major source of cosmic rays up to about $10^{15}-10^{17}$ eV. Propagation of cosmic rays in the galactic medium is known to affect their spectrum and direction distribution [1-13].
The detailed transport properties, however, are not properly understood yet, because of limited knowledge of the interstellar magnetic field at different scales. Recently, there have been improvements in the understanding of the global structure of the galactic magnetic field regular component~\cite{gal1, gal2, gal3}, of the local interstellar magnetic field~\cite{limf} and of the global heliospheric magnetic field~\cite{helio1, helio2}. On the other hand, the scales and spatial distribution of the turbulent component are not well constrained, thus limiting our understanding of the diffusion properties of cosmic rays at high energies as well as of the diffuse galactic gamma ray emission.

Searching for the sources of cosmic rays involves the observation of signatures of cosmic ray interactions at or near their acceleration site via extensive multi-messenger observation campaigns of coincident high energy gamma ray and neutrinos~\cite{taboada}. However, the measurement of cosmic ray spectrum, mass composition~\cite{tamburro} and arrival direction on Earth inevitably probes the properties and spatial distribution of their sources as well as of the long propagation journey through the magnetized medium.

Over the last few decades, observations in the range from several tens GeV to PeV have shown that cosmic rays reach Earth with a small but measurable anisotropy of order 10$^{-4}$-10$^{-3}$. This anisotropy varies with the energy of the cosmic particles, although its topological structure remains substantially unaltered up to about 100 TeV, where it abruptly changes (see~\cite{iclarge10, iclarge12, itlarge13} and references therein). The observed arrival distribution of the cosmic rays is not well described by a simple dipole, but is rather composed of several multipole contributions. Although the dominant terms are dipole and quadrupole harmonics, a non-negligible contribution comes from higher multipole terms, which are related to the observed small-scale anisotropy (see~\cite{icsmall11} and references therein).

Generally speaking, the dipole component of the anisotropy is believed to be a tracer of the cosmic ray source distribution, with the largest contribution from the nearest ones~\cite{blasi2, ptuskin1, pohl, biermann, ptuskin2}. On the other hand a dipolar shape of the cosmic ray spatial distribution can only be reproduced by homogeneous and isotropic diffusion with a constant power law energy dependency. Although the diffusion coefficient is relatively well constrained in the 10-100 GeV energy range by direct isotopic composition observations, its properties at higher energies are fundamentally unknown. A spatial variation of the diffusion coefficient~\cite{tomassetti} or an energy dependency due to particle interactions with pre-existing magnetic turbulence~\cite{blasi3, blasi4} might for instance explain the observed proton and helium spectral hardening (that can also be interpreted as generated at the source~\cite{malkov12})\footnote{It is worth noting that the AMS-02 Collaboration recently showed no indication for a spectral brake in their proton and helium spectra~\cite{ams02}.}. Because the interstellar magnetic field is expected to develop anisotropic turbulence, particle diffusion is anisotropic as well, thus contributing to more complex spatial distributions~\cite{effenberger}.

The existence of significant small-scale features in the cosmic ray arrival distribution might be a signature of non-diffusive processes occurring within the scattering mean free path due to the turbulent realization of the interstellar magnetic field~\cite{giacinti1, biermann}. A heliospheric origin was also proposed to explain the remarkably small angular size and the evidence of a possible spectral hardening~\cite{reconnection1, reconnection2, scattering, drury13}.

The study of cosmic ray anisotropy as a function of energy and mass, and of possible correlations with spectral features, can provide a useful probe into the properties of the local interstellar medium. The understanding of particle transport in astrophysical magnetized media has in fact the potential to unveil signatures of nearby sources of cosmic rays, which may be perhaps too old to emit significant electromagnetic radiation~\cite{erlykin12}.

\section{The IceCube Observatory}
\label{sec:ic}
 
The IceCube Observatory, completed in December 2010, is currently the only km$^3$ scale neutrino telescope collecting data. The detector consists of an array of 5,160 digital optical sensors arranged along 86 cables (or strings) between 1,450 and 2,450 meters below the geographic South Pole, where the deep Antarctic ice is particularly transparent.  IceCube also includes a surface shower array, IceTop, and a dense instrumented core with a lower energy threshold, DeepCore. Decommissioned in 2009, the AMANDA neutrino telescope, consisting of 677 analog optical sensors arranged along 19 strings mostly between 1,500 and 2,000 meters depth, collected data for over a decade. Although designed to detect high energy extra-solar neutrino, IceCube and AMANDA can be used to indirectly study cosmic rays via the high energy muons produced in the interaction of cosmic rays with Earth's atmosphere. The Cherenkov light emitted by the penetrating muons while propagating through the transparent deep ice is recorded by the optical modules. Time and amplitude of the light signals are then used to reconstruct the arrival direction of the muons, and therefore of their parent cosmic rays, and to estimate their energy. Due to their location deep in the ice, IceCube and AMANDA are sensitive to cosmic rays with median particle energy of about 20 TeV.
The surface array IceTop consists of 81 stations, each comprising two tanks of frozen clean water with two optical sensors per tank. IceTop is designed to detect the low-energy, predominantly electromagnetic component of air showers. In particular, the distance between the surface detection stations makes it sensitive to cosmic ray particles with energy in excess of about 300 TeV.

The construction of the IceCube Observatory started in 2004 and physics quality data taking commenced in 2006 with partially constructed instrumentation.

\section{Observations}

\subsection{Energy dependency}
\label{sec:anisoe}

The large number of muon events collected by IceCube (about 10$^{10}$-10$^{11}$ each year, depending on the detector configuration) makes it possible to study the arrival direction distribution of cosmic rays at a level of about 10$^{-5}$. The atmospheric muons share the same direction as the parent cosmic ray particle. Since this study does not require very precisely reconstructed muon directions, all collected and reconstructed events with a median angular resolution of about 3$^{\circ}$ are used. The ice overburden represents a natural energy filter. The energy threshold increases with the angle from the vertical direction because of the longer distance in ice muons have to cross to reach the apparatus. To provide a uniform energy response across the visible southern sky, events were selected using a constant energy cut as a function of zenith angle (see~\cite{iclarge12} for detail). Using a full simulation of the high-energy muon component of cosmic rays air showers we find that the median particle energy of the IceCube data sample is about 20 TeV. By selecting brighter events it is possible to reach higher cosmic ray energies. Fig.~\ref{fig:1mapa} and~\ref{fig:1mapb} show the map of relative intensity, in equatorial coordinates, of the arrival direction of cosmic rays at median energies of about 20 TeV and 400 TeV, respectively. The anisotropy at the lower energy appears to be a smooth continuation of the observations in the northern hemisphere, which is globally consistent in a wide range of energies starting from about 100 GeV. On the other hand, the anisotropy at energies in excess of about 100 TeV has a different structure, as shown in~\cite{eastop, tibet} as well. This might indicate a transition in the contribution of nearby sources of cosmic rays~\cite{blasi2, ptuskin1, ptuskin2}, or simply in the propagation mechanisms that generate the asymmetry~\cite{scattering}. Below about 100 TeV, the global anisotropy is dominated by the dipole and quadrupole components~\cite{icsmall11}. However, at higher energies the main feature appears to be a broad deficit on a relatively flat distribution~\cite{itlarge13}.

The anisotropy above 100 TeV was also measured using the surface array IceTop. Unlike the deep IceCube array, IceTop collects the low energy electromagnetic and muon components of cosmic ray showers. Although still an indirect detection of cosmic rays, the large amount of secondary particles and their lateral distribution provide a better energy resolution. If the electromagnetic component is detected in coincidence with the high energy muon component with IceCube, a good primary mass resolution can be achieved as well. With IceTop it was possible to obtain a sky map of relative cosmic ray intensity at median energies of about 400 TeV and 2 PeV, as shown in Fig.~\ref{fig:1mapc} and~\ref{fig:1mapd}. The latter is the first determination of the anisotropy near the knee of the cosmic ray spectrum, showing the same global structure as at about 400 TeV, but with a deeper deficit~\cite{itlarge13}.

The non-dipolar structure of the anisotropy above a few 100 TeV challenges the current models of cosmic ray diffusion. Whether the strengthening of the deficit region with energy is due to propagation effects from a given source or to the contribution of heavier nuclei at the knee is not clear. In this regard, the study of cosmic ray spectrum and composition in correlation to the anisotropy and vice-versa of the anisotropy for different primary particle masses, would provide an experimental basis to test scenarios of particle transport in the local interstellar medium.

\begin{figure}[t]
\begin{tabular}{c}
\subfloat[]{\includegraphics[width=0.45\textwidth]{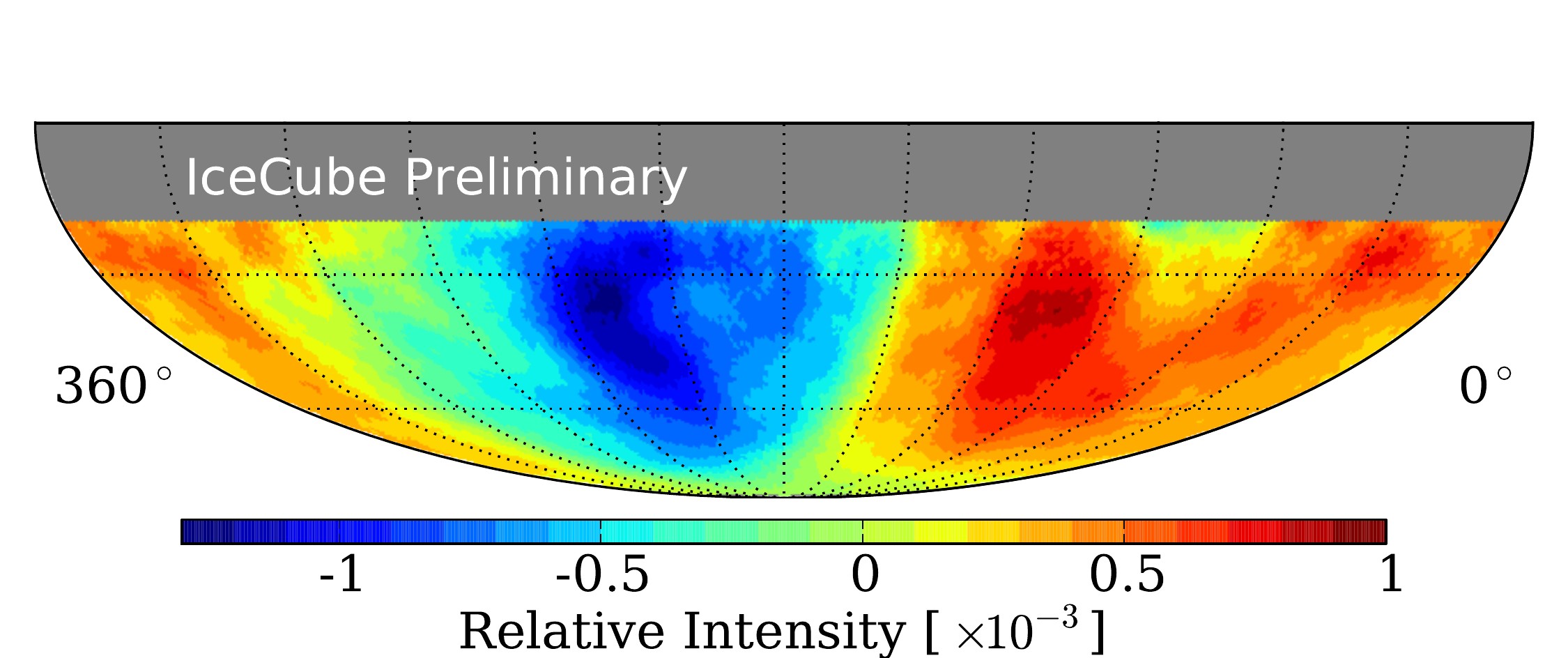} \label{fig:2mapa}} \\
\multicolumn{1}{r}{ } \\
\subfloat[]{\includegraphics[width=0.45\textwidth]{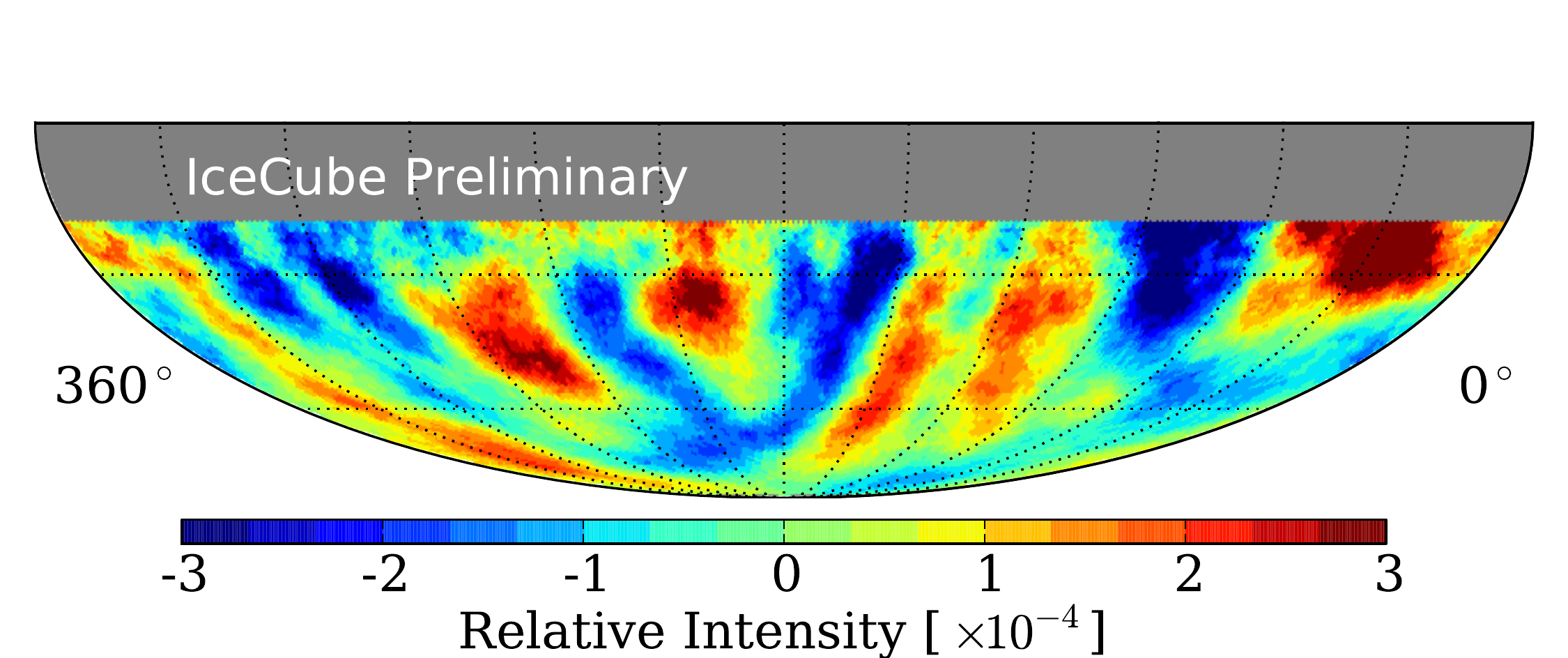} \label{fig:2mapb}} \\
\end{tabular}
\caption{Preliminary 2D maps of relative intensity in equatorial coordinates of the cosmic ray arrival distribution for all data collected by IceCube (at 20 TeV from June 2007 to May 2012)~\cite{ic86}. The maps are shown before (a) and after (b) the dipole and quadrupole subtraction procedure. Maps were smoothed using a circular window of angular radius 5$^{\circ}$.}
\label{fig:2maps}
\end{figure}

\subsection{Angular scale dependency}
\label{sec:anisoas}

Using all data collected by IceCube from June 2007 to May 2012 (about 150 billion events) with a median energy of about 20 TeV, it was possible to derive an angular power spectrum of the relative intensity map~\cite{icsmall11, ic86}. Taking into account the partial sky coverage, the angular spectrum shows that the largest fraction of the total power is in the dipole and quadrupole components of the spherical harmonic expansion (i.e. with index $\ell$ = 1 and 2). However a non-negligible departure from isotropy is also observed at angular scales of about 10$^{\circ}$ to 40$^{\circ}$ ($\ell$ = 5 - 20). The sensitivity to smaller scales is currently limited by event statistics. To highlight the small-scale structure of the cosmic ray arrival distribution, the dipole and quadrupole contributions of the spherical harmonic representations were fit and subtracted from the relative intensity map~\cite{icsmall11, ic86}. Fig.~\ref{fig:2mapa} shows the map with all events collected by IceCube until 2012 at a median energy of 20 TeV, and Fig.~\ref{fig:2mapb} shows the map after dipole and quadrupole fit subtraction.
The small-scale anisotropy structure observed by IceCube at 20 TeV appears to be relatively well correlated to that observed in the northern hemisphere in the TeV energy range [38-43].
Minor differences might originate from the different energy scale and possibly the different mass composition sensitivity of air shower versus muon detectors.

Although astrophysical interpretations of the small-scale cosmic ray features have been proposed (see~\cite{astro1, astro2, astro3}), it is unlikely that 1-10 TeV charged particles are collimated in narrow regions unless the underlying physical processes occur locally, either within the scattering mean free path~\cite{giacinti1, biermann} or due to propagation effects induced by the heliosphere and its boundary with the local interstellar magnetic field~\cite{scattering, drury13}. The possibility that part of the observed localized excess regions are actually extended emissions of TeV gamma rays from the decay of unstable accelerated nuclei was proposed as well~\cite{fargion}.

\begin{figure*}[t]
\includegraphics[width=0.95\textwidth]{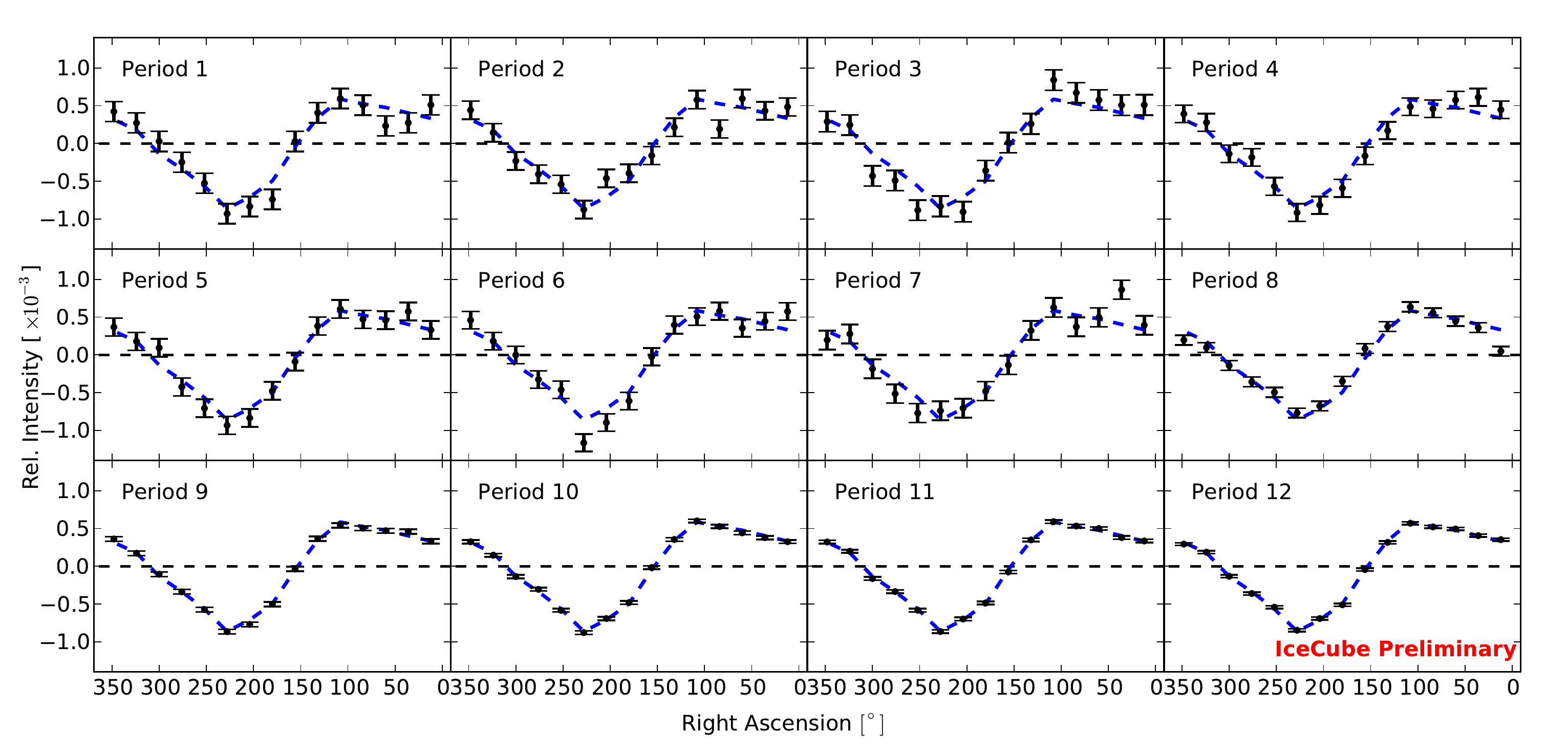}
\caption{Preliminary 1D projections of relative intensity as a function of right ascension for the AMANDA data collected from 2000 to 2006 and IceCube data collected from 2007 to 2011~\cite{amanda}. As a reference the average profile for the entire dataset is shown as a dashed blue line. The uncertainties shown are statistical only.}
\label{fig:3maps}
\end{figure*}

\subsection{Time dependency}
\label{sec:anisot}

It has been extensively documented that galactic cosmic rays below 100 GeV are directly influenced by solar activity at different time scales [48-52].
There is also evidence that the diurnal anisotropy of cosmic rays in the 600 GeV energy range is correlated with solar cycles~\cite{munakata}. However no time dependence was observed by various experiments in the multi-TeV energy range, like for instance~\cite{amenomori}.

Using AMANDA data collected from 2000 to 2006 and IceCube data collected from 2007 to 2012, it was possible to measure the sidereal anisotropy in the energy range of about 10 TeV and compare the yearly observations with the average~\cite{amanda}. As shown in Fig.~\ref{fig:3maps}, the sidereal anisotropy does not exhibit any significant variations over the 12-year time period.












\end{document}